\documentclass[twocolumn]{article}
\usepackage{graphicx}

\begin{document}
\title{Long term behavior of a hypothetical planet in a highly eccentric
  orbit}
\author{R. Nufer\footnote{Im R\"omergarten 1, CH-4106 Therwil, Switzerland},
  W. Baltensperger\footnote{Centro Brasileiro de Pesquisas F\'\i
    sicas, 222\thinspace 90 Rio de Janeiro, Brazil},
  W. Woelfli\footnote{Institute for Particle Physics, ETHZ
    H\"onggerberg, CH-8093 Z\"urich, Switzerland}}
\maketitle

\abstract{For a hypothetical planet on a highly
eccentric orbit, we have calculated the osculating orbital 
parameters and its closest approaches to Earth and Moon over a 
period of 750 kyr. The approaches which are  
close enough to influence the climate of the Earth form a pattern 
comparable to that of the past climatic changes, as recorded in 
deep sea sediments and polar ice cores.} 

\section{Motivation}
The information on Earth's past climate obtained from deep sea sediments
and polar ice cores indicates that the global temperature oscillated with
a small amplitude around a constant mean value for
millions of years until about 3.2 million years ago. From then on the
mean temperature decreased stepwise and the amplitude of the
variations increased. For the last one million years Earth's climate 
was characterized by a distinct 100 kyr periodicity in the advancing
and retreating of the polar ice sheets, known as glacial--interglacial 
cycles. This Ice Age period ended abruptly 11.5 kyr ago. During the
following time interval, the Holocene, the global mean temperature quickly
recovered to a value comparable with that observed 3.2 Myr
ago\ \cite{Tiedemann, Petit}. W\"olfli et al.\ \cite{Woelfli} proposed that 
during all this time an object of planetary size, called Z, existed in a 
highly eccentric orbit, and showed that Z could be the cause of these
changes of Earth's climate. Because of the postulated small perihelion 
distance, Z was heated up by solar radiation and tidal forces so that
it was surrounded by a gas cloud. Whenever the Earth crossed this
cloud, molecules activating a Greenhouse effect were produced in the upper
atmosphere in an amount sufficient to transiently enhance the mean
surface temperature on Earth. Very close flyby events even resulted in
earthquakes and volcanic activities. In extreme cases, a rotation
of the entire Earth relative to its rotation axis occured in response to the
transient strong gravitational interaction. These polar shifts took place 
with a frequency of about one in one million years on the average. The 
first of them was responsible for the major drop in mean temperature, 
whereas the last one terminated Earth's Pleistocene Ice Age 11.5 kyr ago. 
At present, planet Z does not exist any
more. The high eccentricity of Z's orbit corresponds to a small
orbital angular momentum. This could be transferred to one of the
inner planets during a close encounter, so that Z plunged into the
sun. Alternatively, and more likely, Z approached the Earth to less
than the Roche limit during the last polar shift event. In this case
it was split into several parts which lost material at an accelerated
rate because of the reduced escape velocity, so that eventually all of 
these fragments evaporated during the Holocene. 

Here, we describe the method used to calculate the motion of such a
hypothetical object in the presence of the other planets of the solar system, 
and its close 
encounters with the Earth. The calculations neglect possible losses of mass,
orbital energy and angular momentum of Z due to solar irradiation and
tidal effects. We also disregard the disappearance of Z following the last
polar shift event. Consequences of these effects are discussed in
ref.\ \cite{Woelfli}.

\section{The method}
In order to study the motion of an additional planet Z in the
gravitational field of the sun and the other planets, a set of
coupled ordinary differential equations (ODE) has to be integrated
numerically starting at a given time with the known or assumed set of 
orbital parameters of all celestial bodies of interest. For the calculation 
we used the Pascal program {\bf Odeint} which is based on the
Bulirsch-Stoer method. It includes the modified midpoint algorithm
with adaptive stepsize control and Richardson's deferred approach to
the limit\ \cite{Press}. The start values for the known planets were
taken from ref.\ \cite{Ephemeris}, where they are given in barycentric
rectangular coordinates and velocity components referred to the mean
equator and equinox of  J2000.0, the standard epoch, corresponding to 
2000 January 1.5. All values are given in
astronomical units (AU) and AU/day, respectively. To obtain the
heliocentric coordinates referred to the ecliptic, we subtracted the 
coordinates of the sun and rotated the resulting values for all 
planets by 23$^\circ$26'21.448'', the
angle between the mean equatorial plane and the ecliptic of 
J2000.0. The Earth and the Moon were treated separately. The masses 
of the planets are from
ref.\ \cite{System}. For reasons explained in ref.\ \cite{Woelfli} we
assume that Z was a mars--like object with $M_Z=0.11 M_E$ and that its 
hypothetical orbit at the epoch J2000.0 is determined by the 
following heliocentric parameters:
\begin{eqnarray}
\hbox{semi-major axis:}&\hskip0.5cm a& = 0.978\nonumber\\
\hbox{numerical eccentricity:}&\hskip0.5cm e&=0.973\nonumber\\
\hbox{inclination:}&\hskip0.5cm i&=0^\circ\nonumber\\
\hbox{longitude of the perihelion:}&\hskip0.5cm\Omega &= 0^\circ\nonumber\\
\hbox{argument of the perihelion:}&\hskip0.5cm\omega &= 0^\circ\nonumber\\
\hbox{mean anomaly:}&\hskip0.5cm M &= 270.0^\circ\nonumber
\end{eqnarray} 

The calculation was a classical point-mass integration without 
relativistic corrections.
In order to save computing time we ignored the influences from the three 
outermost planets Uranus, Neptune and Pluto, 
and restricted the numerical accuracy per integration step, the tolerance
level $eps$, to a value of $10^{-13}$\ \cite{Press}. 
Several tests were made to check whether these simplifications are
acceptable or not. First of all, we 
evaluated the osculating orbital parameters for the Earth over the past
300 kyr without Z, but for three 
different tolerance levels, $eps = 10^{-13},\; 10^{-15}$ and $10^{-16}$, 
and found that the positions of all planets except 
that of Mercury were reproducible to within less than 100 km. A comparison
of the Earth's eccentricity and 
inclination variations with the corresponding values published by
Berger\ \cite{Berger} and transferred to the 
invariable plane of the solar system by Quinn et al\ \cite{Quinn} also showed 
good agreement. We also determined 
the total angular momentum of the solar system and found a negligible
small linear change at the 
eleventh digit of its value. All calculations were performed with a 133 MHz
PC. Including Z, about 15 h 
were required to cover a time span of 10 kyr with an accuracy of $10^{-13}$.
The sequence of close encounters of Z with the inner planets and the Moon 
amplify the errors so that the calculated result represents a possible 
orbit only. Integrations with Z were performed both, forward (+300 kyr) 
and backward (-450 kyr) in time relative to J2000.0 in order to demonstrate 
that the behavior of the orbital parameters of Z and the Earth were not 
affected by the change in time direction.   

\section{ Results}
Fig.\ 1
\begin{figure*}
\includegraphics[scale=0.4]{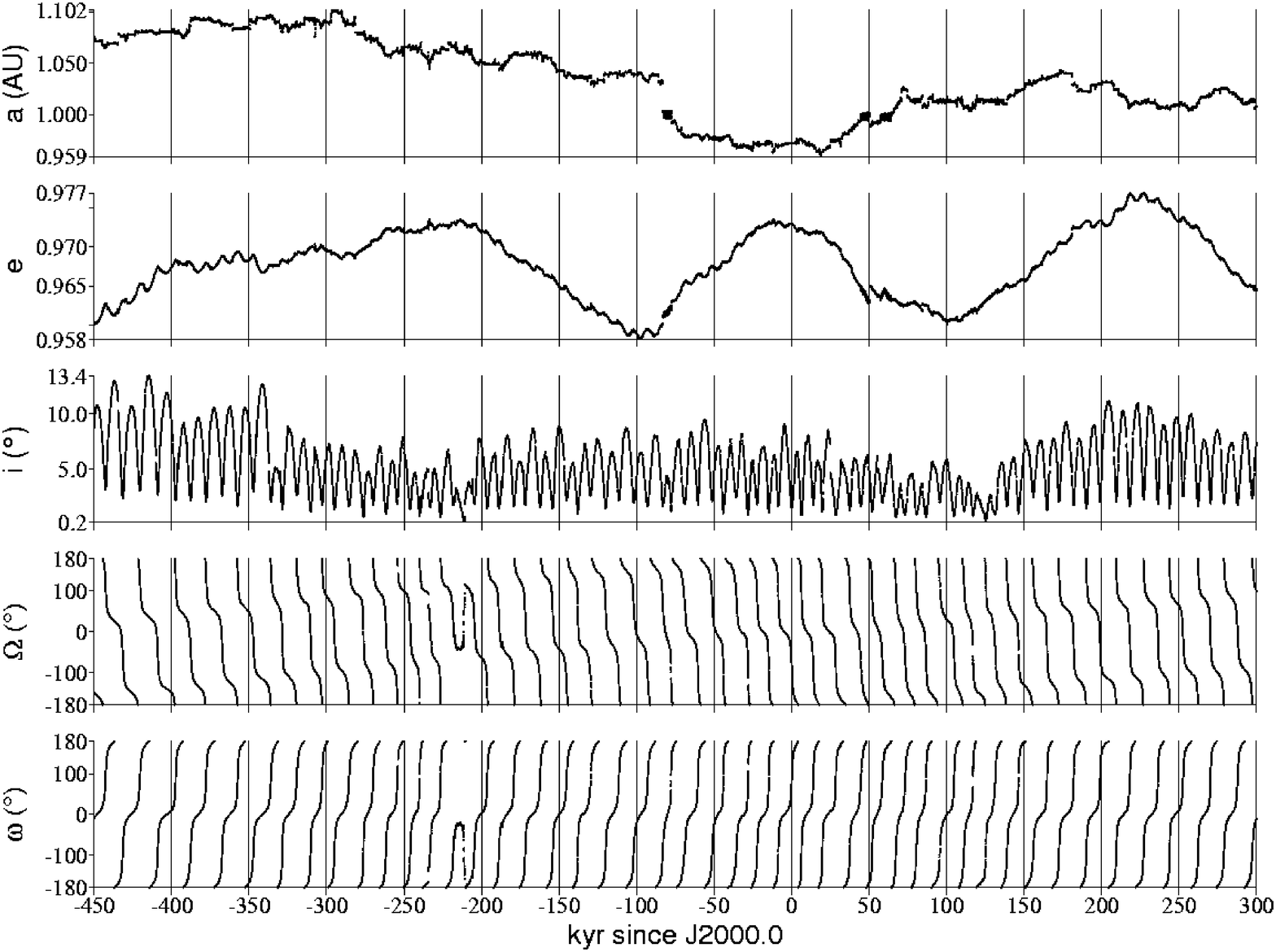}
\caption{Osculating orbital parameters 
($a$, semi-major axis; $e$, eccentricity; $i$, inclination; $\Omega$, 
longitude of the perihelion; $\omega$, argument of the perihelion)
over 750 kyr for the proposed
object of planetary size Z with 
mass $M_Z = 0.11 M_E$. All parameters are evaluated relative to the 
invariable plane which is perpendicular to 
the orbital angular momentum of the solar system. The calculation was
started at $t = 0$ with the orbital 
parameter values listed in section 2. Note the ``high'' frequency of the
inclination.}
\label{nufig1}
\end{figure*}
shows the time dependent variations of the osculating orbital
parameters of Z (semi-major axis, 
eccentricity, inclination, ascending node and argument of the perihelion)
over a time period of 750 kyr.
The inclination is defined here as the angle between Z's orbital plane 
and the invariable plane, which is 
perpendicular to the orbital angular momentum of the solar system. It
is close to the orbital 
plane of Jupiter. The movement of Z is strongly influenced by the inner
planets and to a lesser extent by 
Jupiter. Of interest is the behavior of the inclination, which on average
oscillates with a periodicity of 
about 7 kyr only. This period is more than an order of magnitude shorter
than that of Earth's inclination 
which is essentially determined by Jupiter and, without Z included in 
the calculation, amounts to 100 kyr. The corresponding 
osculating parameters for Earth's orbit are displayed in Fig.\ 2;
\begin{figure*}
\includegraphics[scale=0.4]{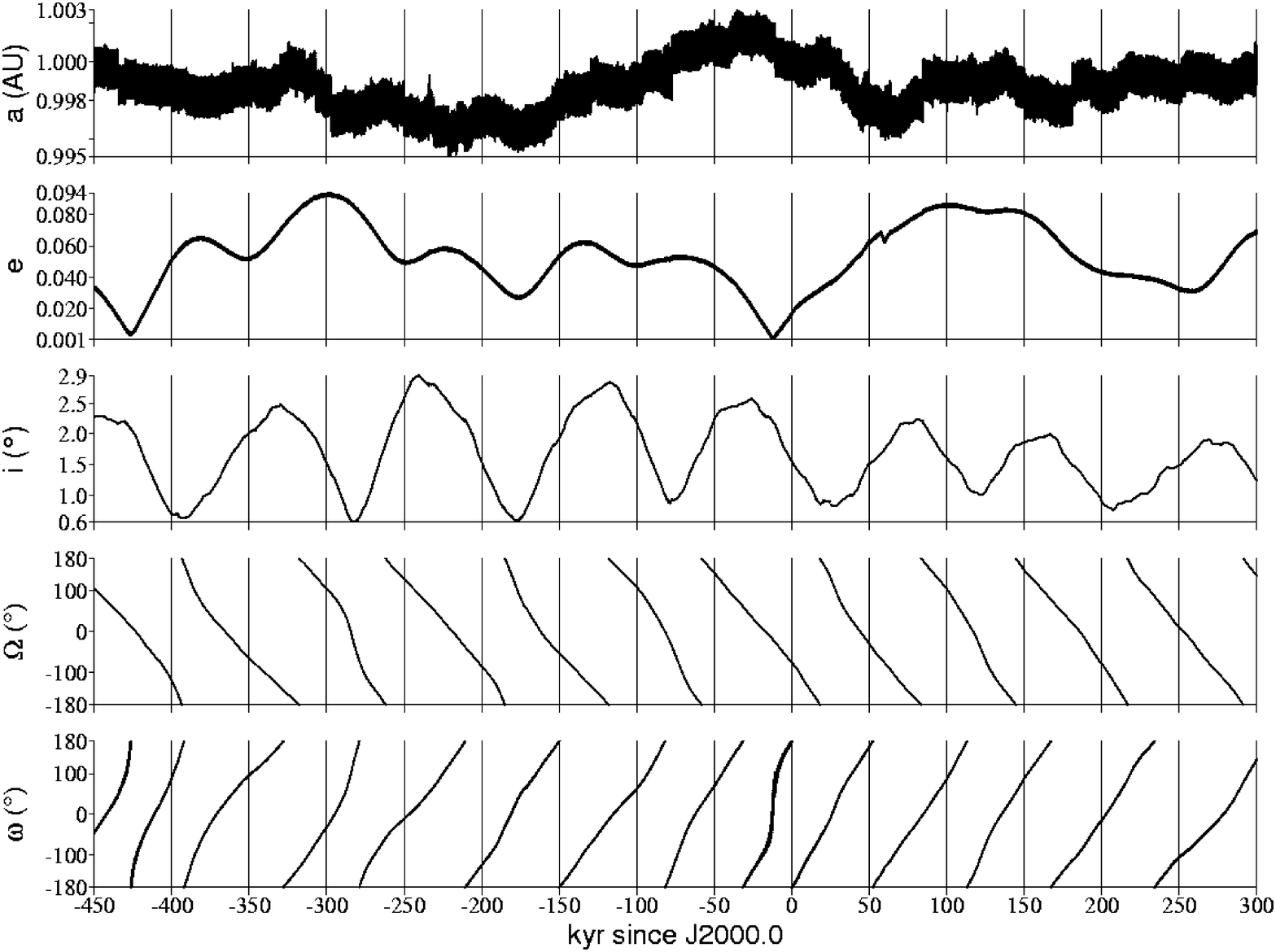}
\caption{Osculating orbital parameters for the Earth
($a$, semi-major axis; $e$, eccentricity; $i$, inclination; $\Omega$, 
longitude of the perihelion; $\omega$, argument of the perihelion)
for the same time
range as in Fig.\ 1. A comparison with the values calculated by 
Berger\ \cite{Berger}
and tranformed by Quinn et al.\ \cite{Quinn} into the
invariable plane suggests that Earth's orbit is only marginally
perturbed by Z.}
\label{nufig2}
\end{figure*}
as mentioned above, they are only 
marginally disturbed by the presence of Z. Sudden jumps in one or several
orbital parameter values of Z 
indicate close encounters with one of the inner planets. 

In order to find
out how often Z approached the 
Earth to distances close enough to influence its climate or even to
provoke a polar shift\ \cite{Woelfli}, we 
determined the distance between Z and the Earth for each integration step
and fitted a parabolic function 
into the values close to the distance of closest approach. The minimum of
this quadratic function was 
then identified with this distance. The upper panel of Fig.\ 3
\begin{figure*}
\includegraphics[width=17.9cm, height=7.6cm]{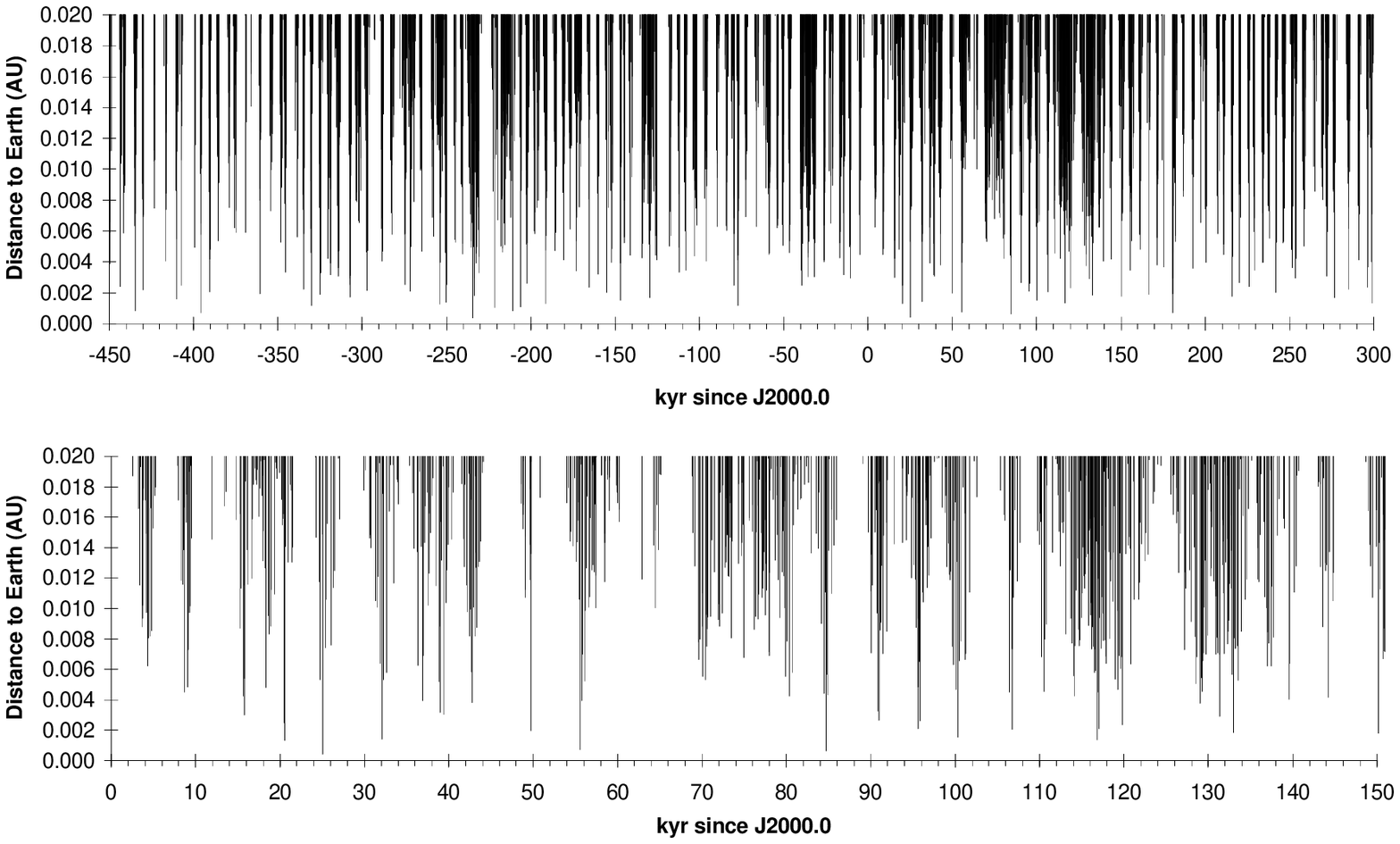}
\caption{Upper panel: Closest approaches of Z to the Earth over 750 kyr.
The plot contains all distances of less 
than 0.02 AU = $3\cdot 10^6$ km, as indicated by the lower endpoints of 
each vertical line. Lower panel: Expanded view 
over 150 kyr, i.e.\ from 0 to + 150 kyr, showing details of the irregular
clustering of these events. This 
structure sensitively depends on the mass of Z and on the initial orbital
parameters selected at $t = 0$.} 
\label{nufig3}
\end{figure*}
shows the
result of this evaluation for the Z--Earth system over the whole time 
range considered here. Plotted are all encounters with distances of 
less than 0.02 AU $= 3\cdot 10^6$ km, as indicated by the endpoints of 
each vertical line. The lower panel of Fig.\ 3 
shows details of the irregular structure of these encounters which are the
result of  the complex time 
dependence of the coordinates of Z and Earth. Not surprisingly,
the encounter frequency is 
enhanced whenever the two inclinations nearly coincide. Fig.\ 4 (left
side)
\begin{figure*}
\includegraphics[width=15.9cm, height=7.6cm]{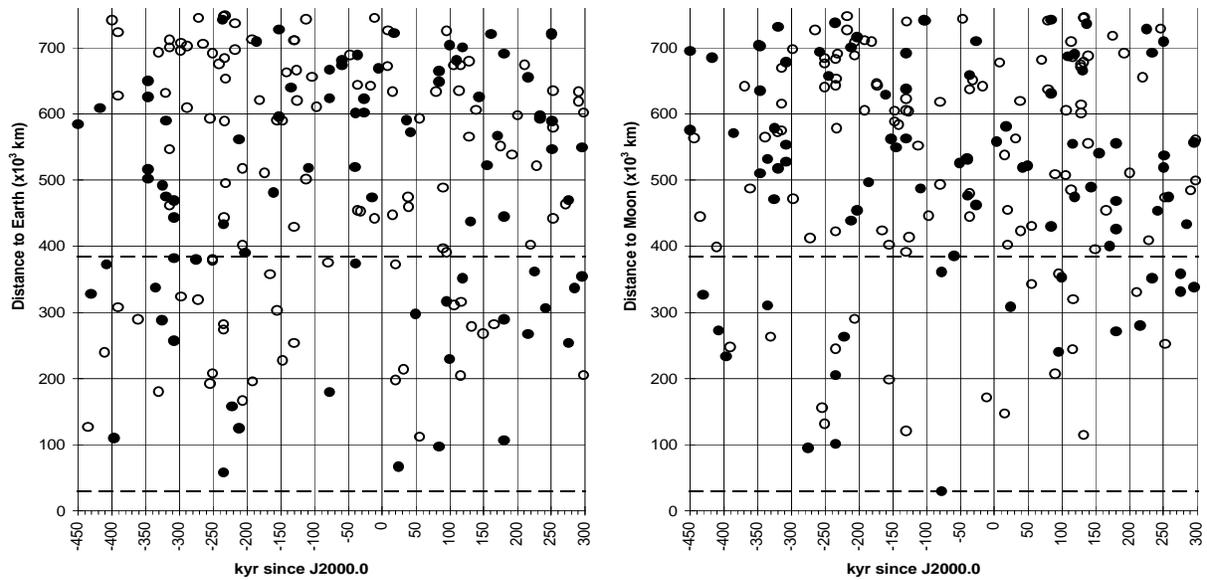}
\caption{Left side: Closest approaches of Z to the Earth below twice the
Moon--Earth distance for the 
same time range. Empty and filled dots correspond to incoming and 
outgoing movements of Z, respectively.
Right side: Analogous representation for the approaches to
the Moon. The two horizontal  
dashed lines in both figures mark Moon's distance (presently
384\thinspace 000 km)
and the critical distance (30\thinspace 000 km), respectively, below
which significant polar shifts on Earth as well as dramatic changes in 
Moon's orbital parameters have to be expected.}
\label{nufig4}
\end{figure*} 
shows that 
Z approaches the Earth many times to less than Moon's distance.  As
explained in ref.\ \cite{Woelfli}  these 
flyby events can excite strong earthquakes and volcanic activities. For
distances smaller than 30'000 km 
such encounters could even result in a rotation of the Earth
by as much as 20$^\circ$  with respect to the direction of the invariant 
angular momentum. Flyby events having distances larger than about
the Moon-Earth distance are harmless in this respect,
but they still may influence Earth's climate. In
ref.\ \cite{Woelfli}  we have shown that Z was 
surrounded by a gas cloud which had an estimated radius of about 
2.8 Mio.\ km at the intersection point with Earth's 
orbit. The interaction of this cloud with Earth's atmosphere produced
Greenhouse gases in 
sufficient amounts to significantly increase the global temperature. 

Close encounters with the Earth also imply close encounters with the 
Moon. These are plotted on the right side of Fig.\ 4.
Since the mass of the Moon is much smaller than that of Z, recoil 
effects are much larger than in the case 
of the Earth and, therefore, may significantly influence the Moon's orbit. In 
fact, Fig.\ 5.
\begin{figure*}
\includegraphics[scale=0.4]{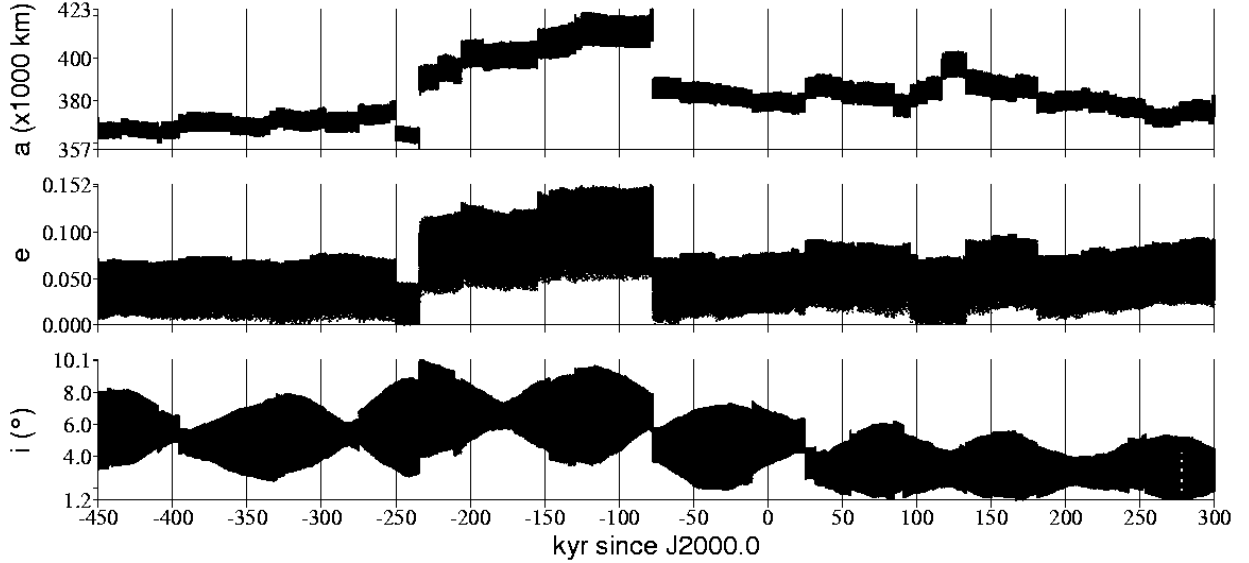}
\caption{Moon's orbital changes over 750 kyr induced by the flyby 
events of Z. Viewed over a sufficiently long time (750 kyr) the net 
effect is small because of the quasi-randomness of these encounters.} 
\label{nufig5}
\end{figure*}
shows events in which the semi-major 
axis $a$ suddenly changes at some given time by up to 9\% relative
to the mean value. Since, according to Kepler's third law, the orbital 
period of the Moon is proportional to $a^{3/2}$, the orbital angular 
frequency $\omega_{orb}$ also changes, so that the apparent rotational
frequency $\Omega = \omega_{rot} - \omega_{orb}$  becomes different 
from zero, assuming that $\omega_{rot}$, the rotational angular
frequency, is not influenced by such an event. In ref.\ \cite{Woelfli} 
we propose that the last close encounter of Z with the Earth took place
only 11.5 kyr ago, so that Moon's orbit could also have changed at that
time. Therefore, the question arises whether the tidal friction on
the Moon is large enough to stop $\Omega$ within the Holocene. In the 
appendix we show that this is likely to be the case.  

\section{Conclusion} 
The calculations presented here show that an object of
planetary size in a highly eccentric 
orbit approaches the Earth with sufficient frequency to influence its
climate and even to produce polar shifts, the last of which terminated 
Earth's Ice Age period, as explained in ref.\ \cite{Woelfli}. Although the 
pattern of these approaches compares well with the observed pattern, we 
have to point out that for various  reasons a one to 
one correspondence between ``theory'' and ``observation'' cannot be 
expected: First, since Z no longer exists at present, we have to 
calculate its long-term behavior on the basis of assumed 
orbital parameters and on estimations of its mass. Second, Z lost 
substantial amounts of mass, orbital 
energy and angular momentum each time it passed through the perihelion.
These effects are neglected in 
our point--mass  model calculation. They may significantly alter
the orbit of Z, and have to be included in an attempt to find out 
whether the disturbances of the orbits of the inner planets due to Z
are within the boundaries set by 
present day observations. In ref.\ 3, we also point out that Z
disappeared from the solar system either 
during the proposed polar shift event 11.5 kyr ago or later on
during the Holocene. A detailed study of the mechanisms responsible 
for this removal is another important task. 

\section*{Appendix: Synchronising Moon's rotation} 
Since a close encounter between Z and the 
Moon can lead to an apparent rotational frequency 
$\Omega  = \omega_{rot} - \omega_{orb}$ of the Moon which no longer 
vanishes on the average, it is important to know how fast tidal 
friction diminishes $\Omega$. The tidal force field is parallel 
to the direction Moon-Earth and has the value
\begin{equation}
                       F = \frac{2 z G M_E R_M}{R^3}.
\end{equation}
$G$ is the gravitational constant, $M_E$ the mass 
of the Earth, $R_M$ the radius of the Moon, $R$ the distance between 
the centers of Earth and Moon, and $z$ the cartesian coordinate in the 
direction Moon--Earth, measured from Moon's center. On Moon's surface
$z=R_M\cos (\gamma )$, where $\gamma$ is the angle between the $z$-axis 
and the direction to the point considered.
Under the influence of $F$
and the gravitational acceleration $g_M$ on Moon's surface, its shape 
will deviate from that 
of a sphere. The deformation is in first order given by
\begin{equation}
     H(\gamma ) = H_0\left(\frac{1}3 + \cos (2\gamma )\right),
\end{equation}
 In equilibrium $H_0$ becomes
\begin{equation}
H_0=\frac{GR_M^2M_E}{2R^3g_M}=6.5 \; \hbox{m}.
\end{equation}
In a time dependent situation elastic tensions reduce the deformation. They
will fade away gradually so that equilibrium is reached, say, with a 
relaxation time $\tau$.  Assuming that the presence of an apparent 
rotation $\Omega$ results in a deformation which lags behind with a phase
difference $\phi = \Omega \tau$, the force field $F$ acting on this 
deformation will exert an angular moment $D$ which tends to
stop the rotation $\Omega$. The integration over Moon's surface yields
\begin{equation}
D = -\frac{\sin (2\phi )}2\;\frac{16\pi}{15}\; 
\frac{G^2 M_E^2 R_M^6 \rho}{R^6 g_M}
\end{equation}
where $\rho$ is the density on Moon's surface. The apparent rotation
$\Omega$ varies in time as
\begin{equation}
   \frac{ d\Omega }{dt} =\frac{D}{\Xi} = - K \frac{\sin (2\phi )}2
\end{equation}
For the inertial moment of the Moon $\Xi$ we use the value of a homogeneous
sphere, $ \Xi  =\frac 25 M_M R_M^2$, and 
set $M_M = \frac{4\pi}3 R_M^3 \rho $. Then $K =\frac{M_E^2GR_M^3}
{M_MR^6}=1.0\cdot 10^{-16}$ s$^{-2}$.
For $\phi = \Omega \tau \ll 1$ the solution of this equation is given
by
\begin{equation}
    \Omega (t) = \Omega (0)\; \hbox{e}^{- K \tau t}   \label{Omega}
\end{equation}
The true relaxation time of this deformation is open to discussion. If, for
example, this time is assumed to be $\tau = 1$ d = 86\thinspace 400 s, 
then the decay constant for $\Omega$ becomes 
\begin{equation}
        \tau_\Omega = \frac{1}{K \tau} = 3500 \hbox{ yr}
\end{equation}
Assuming $\Omega = 0.1 \omega_{rot}$, then
the phase becomes $\phi = 0.1 \omega_{rot}\tau  = 0.023 \hbox{ rad} 
  =  1.3^\circ$.
Thus a rather small phase lag of the tidal deformation is sufficient to
synchronise the lunar rotation within the Holocene. This assumed 
phase lag is smaller than the
value of 2.16$^\circ$ inferred from astronomical data regarding the
actual lunar bulge\ \cite{Gordon}. The fact that according to 
Eq.\ \ref{Omega} \ $\Omega (t)$ never stops, 
is an artefact of a model with a single decay constant. The real 
dynamics also involves slow relaxations, which terminate the 
synchronisation in a finite time. 

\section*{Acknowledgement}
We thank Hans-Ude Nissen for his improvements of the manuscript.


\begin{thebibliography}{99}
\bibitem{Tiedemann}  R. Tiedemann, M. Sarntheim, N.J. Shackleton,
{\it  Astronomic time scale for the Pliocene Atlantic $\delta^{18}$O
and dust flux
record of Ocean Drilling site 659}, Paleocenography, {\bf 9}, (1994)
619-638.

\bibitem{Petit} J.R. Petit, J. Jouzel, D. Raynaud, N.I. Barkov,
  J.-M. Barnola, I. Basile, M. Bender, J. Chappellaz, 
M.  Davis, G. Delaygue, M. Delmotte, V.M. Kozlyakov, M. Legrand, V.Y.
Lipenkov, C. Lorious, 
N.  L. Pépin, C. Ritz, E. Saltzman, M. Stievenard,
{\it Climate and atmospheric history of the past 420'000 years from 
the Vostok ice core, Antartica},
Nature, {\bf 399}, (1999) 429-436.

\bibitem{Woelfli} W. W\"olfli, W. Baltensperger,{ \it  A possible
explanation for Earth's climatic changes in the past few million years},
Notas de Fisica, Centro Brasileiro de Pesquisas Fisicas, Rio de
Janeiro, CBPF-NF-031/99, June 1999. More recent version:
http://xxx.lanl.gov/abs/physics/9907033.

\bibitem{Press} W.H. Press, B.P. Flannery, S.A. Teukolsky, W.T. 
Vetterling,
{\it The Art of Scientific Computing - Numerical Recipes in Pascal},
 Cambridge University Press, Cambridge, 1989.

\bibitem{Ephemeris} JPL Ephemeris DE200/LE200, Magnetic tape from ADC,
private communication, 1992. 

\bibitem{System}{\it System of Astronomical Constants, IAU (1976) and 
Explanatory Supplement to the Astronomical Almanac}, University
Science Books, ed. P.K. Seidelmann, Mill Valley, California, 1992.

\bibitem{Berger} A. Berger,
{\it Accuracy and frequency stability of the Earth's orbital elements
during the Quaternary,
 Milankovitch and Climate}, D. Reidel Publishing Company, 1984, Part.
1, 3--39.

\bibitem{Quinn} T.R. Quinn, S. Tremaine, M. Duncan, {\it A three 
million year integration of the Earth's orbit}, Astronomical Journal, 
{\bf 101}, (1991) 291-355.

\bibitem{Gordon} Gordon J.F. MacDonald, {\it Tidal friction}, Rev.\ of
  Geophysics, {\bf 2}, (1964) 467-541.
\end{thebibliography}
\end{document}